\documentclass[aps,prl,floatfix,twocolumn]{revtex4}
\usepackage{graphicx}

\graphicspath{{figures/}}

\usepackage{amsmath,amssymb}
\usepackage{units,xspace}

\newcommand{\micron}{\ensuremath{\unit{\mu m}}\xspace}

\renewcommand{\vec}[1]{\ensuremath{{\mathbf #1}}\xspace}
\newcommand{\vecr}{\vec{r}}
\newcommand{\avg}[1]{\left< #1 \right>}
\newcommand{\abs}[1]{\left| #1 \right|}

\begin{document}
  
\title{Colloidal Electrostatic Interactions Near a Conducting Surface}

\author{Marco Polin}
\author{David G. Grier}

\affiliation{Department of Physics and Center for Soft Matter Physics,
  New York University, 4 Washington Place, New York, NY 10003}

\author{Yilong Han}

\affiliation{Department of Physics and Astronomy, University of Pennsylvania,
  209 South 33rd St., Philadelphia, PA 19104}

\date{\today}

\begin{abstract}
Charged-stabilized colloidal spheres dispersed in deionized
water are supposed to repel each other.
Instead, artifact-corrected video microscopy measurements reveal
an anomalous long-ranged like-charge attraction in the interparticle
pair potential
when the spheres are confined to a layer by even a single charged
glass surface.
These attractions can be masked by electrostatic repulsions
at low ionic strengths.
Coating the bounding surfaces with a conducting gold layer
suppresses the attraction.
These observations suggest a possible mechanism for
confinement-induced attractions.
\end{abstract}

\maketitle

Charge-stabilized colloidal spheres carrying the same sign charge and
dispersed in deionized water are predicted to repel each other
by the mean-field theory of macroionic interactions.
This reasonable prediction has been repeatedly challenged
by experimental observations that have been interpreted as
evidence for like-charge colloidal attractions.
Several methods have been introduced for probing these interactions
directly,
the most sensitive of which are based on digital video microscopy
measurements of individual spheres' trajectories.
When applied to dispersions of highly charged micrometer-scale spheres
in deionized water, these
techniques have indeed revealed anomalous long-ranged like-charge attractions
in the effective pair potential \cite{kepler94,crocker96a,carbajaltinoco96,han03a}, 
but only when the spheres are rigidly confined to thin layers by charged glass 
surfaces \cite{crocker96a,behrens01a,han03a}.
Isolated pairs of spheres, by contrast, are found to repel each other
in excellent agreement with standard theoretical predictions
\cite{crocker94,vondermassen94,crocker96a}.

\begin{figure}[t!]
  \centering
  \includegraphics[width=0.9\columnwidth]{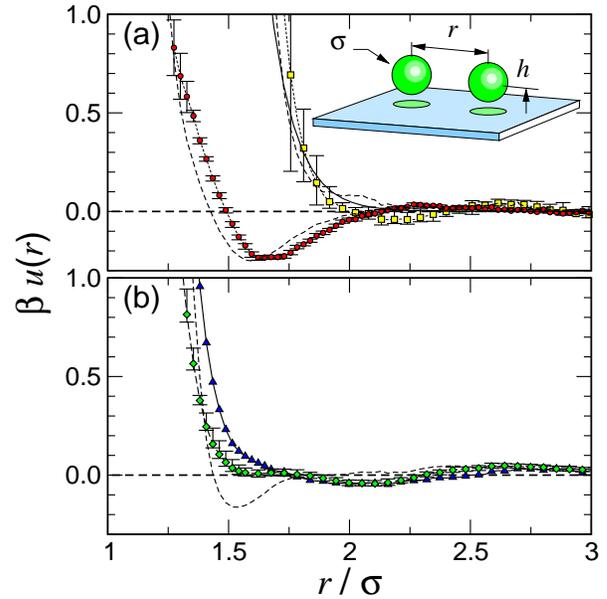} 
  \caption{Measured pair potentials $u(r)$ for $\sigma = 1.58~\micron$ diameter silica 
    spheres near a single wall ($H \simeq 200~\micron$).
    (a) Glass surface with $\kappa^{-1} = 180 \pm 10~\unit{nm}$
    (squares), showing monotonic repulsion,
    and $\kappa^{-1} \approx 60~\unit{nm}$ (circles), displaying long-ranged attraction.
    Inset: schematic of the geometry.
    The solid curve is a fit to Eq.~\protect\ref{eq:dlvo} with
    $Z = 6500 \pm 1000$.
    (b) Single gold-coated surface (diamonds) with $\kappa^{-1} \approx 100~\unit{nm}$.
    Potentials obtained with the HNC closure are plotted.  Those
    obtained with the PY closure are indistinguishable.
    Dashed curves are uncorrected results.  Correcting for imaging
    artifacts removes an apparent minimum in (b), but not in (a).
    The corrected pair potential also is purely repulsive for a thin cell
    ($H = 15~\micron$) with two gold-coated surfaces (triangles).
  }
  \label{fig:urgrgold}
\end{figure}

These observations have proved controversial, both because 
the effect has resisted explanation,
and also because a host of experimental
artifacts might have mimicked attractions in the published
measurements.
In the first place, like-charged colloidal
attractions are inconsistent with the Poisson-Boltzmann mean-field theory
for macroionic interactions
\cite{sader99,neu99,trizac99,sader00}. 
The observed attractions are too strong and long-ranged to be accounted for by van der
Waals interactions \cite{kepler94,pailthorpe82}.
Efforts to explain the
observations on the basis of other non-mean-field mechanisms have proved
inconclusive or unsuccessful \cite{gopinathan02}.
Simulations on smaller, simpler systems also have failed
to reproduce the experimental observations.

Given the effect's apparent subtlety and the challenge it poses to
basic notions in the field, ruling out experimental
artifacts takes on particular importance.
The introduction of thermodynamically self-consistent 
analytical tools \cite{han03a,han04,han05a} ensures that
colloidal pair potentials extracted from trajectory data are free from
such spurious effects as
nonequilibrium kinematic coupling \cite{squires00} 
and uncorrected many-body correlations
\cite{brunner02}.
These tests, however, cannot account for systematic errors in the
trajectory data themselves.
Imaging artifacts that
displace colloidal spheres' apparent centroids
in digital microscopy images
\cite{baumgartl05,baumgartl05a} recently have been shown to mimic
anomalous attractions at least under some circumstances \cite{baumgartl05}.
Their discovery has therefore cast doubt on all previous reports
of like-charge colloidal attractions.

This Letter presents experimental evidence 
for anomalous confinement-induced like-charge colloidal attractions
obtained with thermodynamically self-consistent distortion-corrected
video microscopy measurements.
Having confirmed the appearance of anomalous confinement-induced
like-charge attractions under conditions comparable to 
those described previously,
we further explore the role of surface properties
in modifying colloidal electrostatic interactions.
Previous studies have focused on the influence of pairs of
closely spaced charged
glass surfaces on the interactions between
micrometer-scale spheres composed of polystyrene sulfate
\cite{kepler94,crocker96a,carbajaltinoco96,han03a}, 
silica \cite{han03a,han04,han05a} 
and polymethyl methacrylate (PMMA) \cite{crocker96phd} dispersed in
clean water.
One exception is a study of PMMA spheres confined by neutral
elastomer surfaces, in which the appearance of long-ranged attractions
was ascribed to
capillary forces \cite{cui02}.
We demonstrate that attractions between colloidal silica spheres in
equilibrium can be induced even by a single charged glass wall, but 
are not by gold-coated surfaces
under otherwise identical experimental conditions.
These results are summarized in Fig.~\ref{fig:urgrgold}.

Our samples consist of uniform silica spheres
$\sigma = 1.58 \pm 0.06~\micron$ in diameter (Duke Scientific Lot 24169) 
dispersed 
in deionized water and loaded into hermetically sealed sample volumes
formed by bonding the edges of glass \#1.5 coverslips to the surfaces of glass
microscope slides separated by $H = 200~\micron$.
Access to the sample volume is provided by glass tubes bonded to holes 
drilled through the slides.
Filling the tubes with mixed bed ion exchange resin maintains the ionic strength below
$5~\unit{\mu M}$, corresponding to a Debye-H\"uckel screening length of
$\kappa^{-1} \approx 200~\unit{nm}$.
Removing the ion exchange resin and allowing the dispersions to
equilibrate with air reduces the screening length to $\kappa^{-1}
\approx 60~\unit{nm}$.
To investigate the influence of the bounding surfaces' properties on
confined colloids' interactions, we coated the inner surfaces
of some volumes with 10~\unit{nm} gold films on
10~\unit{nm} titanium wetting layers before assembly.
These metallic films have a resistivity of $50~\unit{\Omega/\Box}$
and are optically transparent.
Sealed samples were mounted on the stage of a 
Zeiss Axiovert 100 STV microscope where they
were allowed to equilibrate at room temperatures ($T = 297 \pm 1~\unit{K}$).
The dense silica spheres rapidly sediment into a monolayer with their
centers about 0.9~\micron above the lower wall and
with out-of-plane fluctuations smaller than 300~\unit{nm} 
\cite{behrens01a,han03a}. 
The more distant wall is far enough from the sedimented spheres at the upper end of
this range that we ascribe any anomalous effects to confinement
by a single wall.

The bright-field imaging system provides a magnification of $212~\unit{nm/pixel}$
on a Hitachi TI-11A monochrome charge coupled device (CCD) camera.
The resulting video stream is recorded and digitized into 10 deinterlaced video
fields per second, each of which is analyzed \cite{crocker96} to
recover the single spheres' centroids, typically identified to within 1/10 pixel.

\begin{figure}[htbp]
  \centering
  \includegraphics[width=\columnwidth]{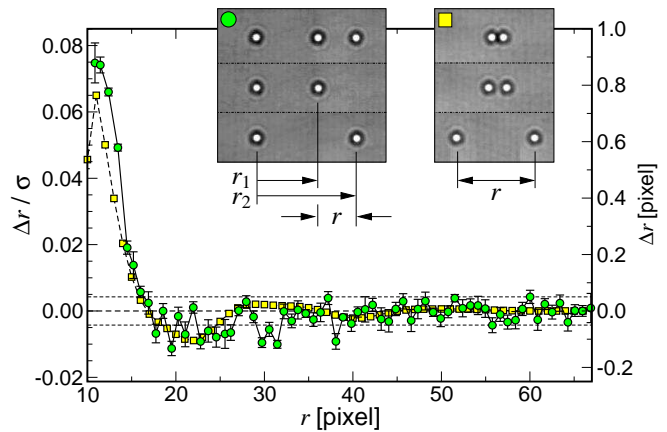}
  \caption{Measurement of the imaging artifact due to overlap of
  single-sphere light scattering patterns.
  Circles represent data obtained with holographic optical tweezer
  measurements and squares with computer-generated composite images.}
  \label{fig:artifact}
\end{figure}

Indivdual spheres' images extend beyond their bright centers
with alternating dark and bright diffraction fringes, which are far-field 
projections of the micrometer-scale particles' light scattering patterns
\cite{bohren83,ovryn00}.
When two particles are close enough for their visible scattering
patterns to overlap, the distortions induced in their individual
images cause systematic deviations in the
spheres' measured separations \cite{baumgartl05}. 
These non-monotonic separation-dependent deviations can affect the qualitative form of the
inter-particle pair potential estimated from measured particle
positions, in some cases creating the appearance of long-ranged
attractions when none exist \cite{baumgartl05}.
Fortunately, this effect can be measured, and the measurements
used to correct the derived results \cite{baumgartl05}.

Figure~\ref{fig:artifact} shows two complementary
approaches for measuring this imaging artifact, one of which can be
applied \emph{a posteriori} to archival data.
In the first, holographic optical
tweezers \cite{dufresne98,polin05} are used to trap three 
colloidal spheres in a line.  
One sphere is held far enough from the other two to avoid any
distortions, and is used as a reference.
The two other traps are set at a fixed separation and filled
with particles, first one at a time, and then together.
The apparent positions, $r_1$ and $r_2$, of these spheres relative
to the reference particle are measured \cite{crocker96} both with
and without the neighbor.
The difference between the spheres' true separation,
$r = r_2 - r_1$ ,and the apparent separation
when both traps are filled is a measurement of the
distortion $\Delta r$, which is plotted as a function of 
true separation in Fig.~\ref{fig:artifact}.  
As previously reported \cite{baumgartl05}, these deviations
substantially exceed the quoted error bound for 
single-particle tracking \cite{crocker96} at separations
relevant for colloidal interaction measurements.

The optical trapping method must be performed \emph{in situ} and thus cannot be
applied to the data in previously published studies
\cite{kepler94,crocker96a,carbajaltinoco96,han03a,han04,han05a}.
Equivalent results can be obtained by analyzing images
of single isolated spheres, and so can be applied after the fact.
Digitally translating a copy of such an image
and superimposing it on the original
yields a composite image of known separation, equivalent to the
incoherent superposition of two single-particle scattered fields.
Examples are shown in the second inset to Fig.~\ref{fig:artifact}.
The apparent separation in each composite can be measured
as before,
and the difference between this and the known
separation is a measurement of the separation-dependent deviation.  
As shown in Fig.~\ref{fig:artifact}, the results agree well
for the spheres in our studies.
The successful comparison with the direct optical tweezer measurement
confirms this method's accuracy, and it is used in the following
measurements.

We estimate a confined dispersion's effective pair potential by
amassing data on equilibrium pair separations over 
periods of roughly one hour \cite{behrens01a}, with temperature
fluctuations maintained below $\pm 1^\circ\unit{C}$.
Time-resolved particle distribution data then are distilled into the
radial distribution function,
\begin{equation}
  \label{eq:gr}
  g(r) = \frac{1}{n^2} \, \avg{\rho(\vec{r}^\prime + \vec{r}, t) \, \rho(\vec{r},t)}
\end{equation}
where $n = N/A$ is the areal density of $N = \avg{N(t)}$ particles in 
the field of view of area $A$, $\rho(\vec{r},t)$ is the instantaneous
particle distribution at time $t$, and the averages are over angles and time.
To assess the influence of separation-dependent imaging artifacts we analyzed both $g(r)$
and the result obtained by approximately correcting for systematic separation distortions,
$\tilde g(r) = g(r + \Delta r) \, (1 + \frac{d}{dr} \, \Delta r)$.
Our samples are dilute enough that higher-order many-body distortions
have a negligibly small effect.

Estimates for the effective
pair potential can be extracted from $g(r)$ and $\tilde g(r)$
in either the hypernetted-chain (HNC)
or Percus-Yevick (PY)
approximations as \cite{chan77,behrens01a}
\begin{equation}
  \label{eq:oz}
  \beta u(r) = - \ln g(r) + 
  \begin{cases}
    n \, I(r) & \text{(HNC)}\\
    \ln (1 + n \, I(r)) & \text{(PY)} 
  \end{cases}
\end{equation}
where $\beta^{-1} = k_B T$ is the thermal energy scale, and
where the convolution integral
\begin{equation}
  \label{eq:ir}
  I(r) = \int_A \left[ g(r') - 1 - n I(r') \right]
  \left[ g(\abs{{\vecr'- \vecr}})-1  \right] d^2r'
\end{equation}
is solved iteratively, starting with $I(r) = 0$.
The results then can be tested for thermodynamic
self-consistency using previously reported
methods \cite{han04,han05a}.

Typical results for silica spheres near glass and gold-coated
surfaces are plotted in Fig.~\ref{fig:urgrgold}(a) and (b) respectively.
Uncorrected data are plotted as dashed curves in
Fig.~\ref{fig:urgrgold},
and corrected results as discrete points.

Imaging artifacts have little influence on the data in
Fig.~\ref{fig:urgrgold}(a).
Colloidal silica spheres hovering over a single charged glass
surface at low ionic strength repel each other, in agreement
with previously reported measurements \cite{behrens01a,han03a,han04}.
The purely monotonic pair potentials obtained under these conditions
also agree with predictions of mean-field theory \cite{russel89},
\begin{equation}
  \label{eq:dlvo}
  \beta u(r) = {Z^\ast}^2 \, \lambda_B \, \frac{\exp(-\kappa r)}{r},
\end{equation}
where 
$Z^\ast = Z \, \exp(\kappa \sigma/2)/(1 + \kappa \sigma/2)$,
is the spheres' effective charge number,
where the Bjerrum length, $\lambda_B = \beta e^2/(4\pi \epsilon_0 \epsilon) = 0.717~\unit{nm}$, 
sets thermal range for electrostatic interactions in a
medium of dielectric constant $\epsilon$, 
and where the Debye-H\"uckel screening length $\kappa^{-1}$ sets the
range over which electrostatic interactions are screened
by the concentration $c$ of
(monovalent) ions.
The fit charge number $Z = 6500 \pm 1000$ is consistent with previous
measurements
on similar spheres \cite{behrens01a}, even after correcting for
imaging artifacts.
The fit screening length $\kappa^{-1} = 180 \pm
10~\unit{nm}$
corresponds to  $c = \kappa^2/(4 \pi \lambda_B) = 5.7~\unit{\mu M}$.

Qualitatively different results are obtained when the ionic strength
is allowed to increase to roughly $50~\unit{\mu M}$
(Fig.~\ref{fig:urgrgold}(a, circles)).
Under these circumstances, the measured pair potential has a minimum
$0.2~k_B T$ deep at a center-to-center separation of 2.7~\micron 
($1.7 \, \sigma$), even
after correcting for imaging artifacts.
Unlike previous reports of attractions induced by pairs of closely
spaced glass walls
\cite{kepler94,crocker96a,carbajaltinoco96,han03a,han04}, the data
in Fig.~\ref{fig:urgrgold}(a) suggest that
even a single surface suffices to induce a
long-ranged attraction between nearby pairs of spheres.
This is qualitatively consistent with the observation that the
repulsion between a pair of charged colloidal spheres
is diminished by introducing a third sphere
into their vicinity \cite{russ05}, with the charged wall in our
experiment playing the role of the third sphere.

Reanalyzing data from Refs.~\cite{han03a} and \cite{han04}
with the \emph{a posteriori} correction method also confirms the
presence of anomalous attractions in more tightly
confined charge-stabilized dispersions.
Furthermore, the increasing prominence of the attractive component of
the interaction with decreasing wall separation \cite{han03a}
appears to have been correctly interpreted as
intrinsic properties of the inter-particle pair potential.
The attraction's apparent dependence on ionic strength also recalls
a similar trend observed in the earliest report of the effect \cite{kepler94}.

Imaging artifacts play a more striking role for the data in
Fig.~\ref{fig:urgrgold}(b, diamonds).
In this case, an apparent minimum in the uncorrected pair potential
disappears entirely after accounting for overlapping images.
The absence of an attraction is noteworthy because
the data in Figs.~\ref{fig:urgrgold}(a,circles) and (b,diamonds) were obtained
under comparable conditions of ionic strength, the only difference
being the gold coating on the bounding surface in (b).
A comparable result is obtained for a thin sample ($H = 15~\micron$)
where both surfaces are gold-coated
(Fig.~\ref{fig:urgrgold}(b,triangles)).
Reducing the inter-wall separation in bare glass cells generally has
been found to strengthen wall-induced attractions \cite{crocker96a,han03a}.
These observations therefore suggest that
nominally uncharged gold surfaces do not induce anomalous
attractions under conditions for which charged glass surfaces
do.

Confirming the importance of surface charge in confinement-induced
attractions helps to narrow the possible explanations for the effect.
A growing body of calculations show that
large, highly charged colloidal spheres can induce 
non-monotonic correlations in the
distribution of simple ions, not captured by mean-field theory, even
in 1:1 electrolytes \cite{hastings78,lee97,carbajaltinoco02}, and that
these can mediate effective inter-particle attractions.
Related calculations suggesting that bounding charged walls also induce
such correlations between \emph{pairs} of spheres
\cite{goulding98,goulding99a} have proved controversial 
\cite{trizac99,mateescu01,trizac01}, although they appear to be
consistent with numerical simulations \cite{allahyarov99}.
These excess correlations correspond to departures from local
electroneutrality in the electrolyte surrounding the macroions.
Under some conditions, they
tend to inject extra counterions 
between pairs of spheres, thereby inducing an effective attraction.
Modeling this diffuse space charge as a point charge, $q$,
at the midpoint between the spheres' centers 
suggests an effective pair potential
\cite{grier04f}
\begin{equation}
  \label{eq:2dlvo}
  \beta u(r) = Z^\ast \lambda_B \,
  \left(Z^\ast \, \frac{\exp(-\kappa r)}{r} 
    + 4 q \, \frac{\exp(- \kappa r/2)}{r} \right),
\end{equation}
which agrees well with measured results, such as those in
Fig.~\ref{fig:urgrgold}.
The solid curves in Fig.~\ref{fig:urgrgold}(a) are both
consistent with $q = 10 \pm 5$.  Those in Fig.~\ref{fig:urgrgold}(b)
are consistent with $q = 0$.
In the latter case, the nominally uncharged walls would have no excess
counterions to contribute.

This semi-heuristic space charge model demonstrates that
correlation-induced attractions beyond the mean-field approximation
can be masked by long-ranged electrostatic repulsions in systems at
very low ionic strength.
Less efficient deionization at small wall separation
might then explain why wall-induced attractions have been observed \cite{han03a} for wall
spacings as large as $H = 30~\micron$, but not previously for 
$H \gtrsim 200~\micron$
\cite{behrens01a,han03a}.

This work was supported by the donors of the Petroleum Research Fund of the American Chemical Society.


\begin{thebibliography}{39}
\expandafter\ifx\csname natexlab\endcsname\relax\def\natexlab#1{#1}\fi
\expandafter\ifx\csname bibnamefont\endcsname\relax
  \def\bibnamefont#1{#1}\fi
\expandafter\ifx\csname bibfnamefont\endcsname\relax
  \def\bibfnamefont#1{#1}\fi
\expandafter\ifx\csname citenamefont\endcsname\relax
  \def\citenamefont#1{#1}\fi
\expandafter\ifx\csname url\endcsname\relax
  \def\url#1{\texttt{#1}}\fi
\expandafter\ifx\csname urlprefix\endcsname\relax\def\urlprefix{URL }\fi
\providecommand{\bibinfo}[2]{#2}
\providecommand{\eprint}[2][]{\url{#2}}

\bibitem[{\citenamefont{Kepler and Fraden}(1994)}]{kepler94}
\bibinfo{author}{\bibfnamefont{G.~M.} \bibnamefont{Kepler}} \bibnamefont{and}
  \bibinfo{author}{\bibfnamefont{S.}~\bibnamefont{Fraden}},
  \bibinfo{journal}{Phys. Rev. Lett.} \textbf{\bibinfo{volume}{73}},
  \bibinfo{pages}{356} (\bibinfo{year}{1994}).

\bibitem[{\citenamefont{Crocker and Grier}(1996{\natexlab{a}})}]{crocker96a}
\bibinfo{author}{\bibfnamefont{J.~C.} \bibnamefont{Crocker}} \bibnamefont{and}
  \bibinfo{author}{\bibfnamefont{D.~G.} \bibnamefont{Grier}},
  \bibinfo{journal}{Phys. Rev. Lett.} \textbf{\bibinfo{volume}{77}},
  \bibinfo{pages}{1897} (\bibinfo{year}{1996}{\natexlab{a}}).

\bibitem[{\citenamefont{Carbajal-Tinoco
  et~al.}(1996)\citenamefont{Carbajal-Tinoco, Castro-Rom\'{a}n, and
  Arauz-Lara}}]{carbajaltinoco96}
\bibinfo{author}{\bibfnamefont{M.~D.} \bibnamefont{Carbajal-Tinoco}},
  \bibinfo{author}{\bibfnamefont{F.}~\bibnamefont{Castro-Rom\'{a}n}},
  \bibnamefont{and} \bibinfo{author}{\bibfnamefont{J.~L.}
  \bibnamefont{Arauz-Lara}}, \bibinfo{journal}{Phys. Rev. E}
  \textbf{\bibinfo{volume}{53}}, \bibinfo{pages}{3745} (\bibinfo{year}{1996}).

\bibitem[{\citenamefont{Han and Grier}(2003)}]{han03a}
\bibinfo{author}{\bibfnamefont{Y.}~\bibnamefont{Han}} \bibnamefont{and}
  \bibinfo{author}{\bibfnamefont{D.~G.} \bibnamefont{Grier}},
  \bibinfo{journal}{Phys. Rev. Lett.} \textbf{\bibinfo{volume}{91}},
  \bibinfo{pages}{038302} (\bibinfo{year}{2003}).

\bibitem[{\citenamefont{Behrens and Grier}(2001)}]{behrens01a}
\bibinfo{author}{\bibfnamefont{S.~H.} \bibnamefont{Behrens}} \bibnamefont{and}
  \bibinfo{author}{\bibfnamefont{D.~G.} \bibnamefont{Grier}},
  \bibinfo{journal}{Phys. Rev. E} \textbf{\bibinfo{volume}{64}},
  \bibinfo{pages}{050401(R)} (\bibinfo{year}{2001}).

\bibitem[{\citenamefont{Crocker and Grier}(1994)}]{crocker94}
\bibinfo{author}{\bibfnamefont{J.~C.} \bibnamefont{Crocker}} \bibnamefont{and}
  \bibinfo{author}{\bibfnamefont{D.~G.} \bibnamefont{Grier}},
  \bibinfo{journal}{Phys. Rev. Lett.} \textbf{\bibinfo{volume}{73}},
  \bibinfo{pages}{352} (\bibinfo{year}{1994}).

\bibitem[{\citenamefont{Vondermassen et~al.}(1994)\citenamefont{Vondermassen,
  Bongers, Mueller, and Versmold}}]{vondermassen94}
\bibinfo{author}{\bibfnamefont{K.}~\bibnamefont{Vondermassen}},
  \bibinfo{author}{\bibfnamefont{J.}~\bibnamefont{Bongers}},
  \bibinfo{author}{\bibfnamefont{A.}~\bibnamefont{Mueller}}, \bibnamefont{and}
  \bibinfo{author}{\bibfnamefont{H.}~\bibnamefont{Versmold}},
  \bibinfo{journal}{Langmuir} \textbf{\bibinfo{volume}{10}},
  \bibinfo{pages}{1351} (\bibinfo{year}{1994}).

\bibitem[{\citenamefont{Sader and Chan}(1999)}]{sader99}
\bibinfo{author}{\bibfnamefont{J.~E.} \bibnamefont{Sader}} \bibnamefont{and}
  \bibinfo{author}{\bibfnamefont{D.~Y.~C.} \bibnamefont{Chan}},
  \bibinfo{journal}{J. Colloid Interface Sci.} \textbf{\bibinfo{volume}{213}},
  \bibinfo{pages}{268} (\bibinfo{year}{1999}).

\bibitem[{\citenamefont{Neu}(1999)}]{neu99}
\bibinfo{author}{\bibfnamefont{J.~C.} \bibnamefont{Neu}},
  \bibinfo{journal}{Phys. Rev. Lett.} \textbf{\bibinfo{volume}{82}},
  \bibinfo{pages}{1072} (\bibinfo{year}{1999}).

\bibitem[{\citenamefont{Trizac and Raimbault}(1999)}]{trizac99}
\bibinfo{author}{\bibfnamefont{E.}~\bibnamefont{Trizac}} \bibnamefont{and}
  \bibinfo{author}{\bibfnamefont{J.-L.} \bibnamefont{Raimbault}},
  \bibinfo{journal}{Phys. Rev. E} \textbf{\bibinfo{volume}{60}},
  \bibinfo{pages}{6530} (\bibinfo{year}{1999}).

\bibitem[{\citenamefont{Sader and Chan}(2000)}]{sader00}
\bibinfo{author}{\bibfnamefont{J.~E.} \bibnamefont{Sader}} \bibnamefont{and}
  \bibinfo{author}{\bibfnamefont{D.~Y.~C.} \bibnamefont{Chan}},
  \bibinfo{journal}{Langmuir} \textbf{\bibinfo{volume}{16}},
  \bibinfo{pages}{324} (\bibinfo{year}{2000}).

\bibitem[{\citenamefont{Pailthorpe and Russel}(1982)}]{pailthorpe82}
\bibinfo{author}{\bibfnamefont{B.~A.} \bibnamefont{Pailthorpe}}
  \bibnamefont{and} \bibinfo{author}{\bibfnamefont{W.~B.}
  \bibnamefont{Russel}}, \bibinfo{journal}{J. Colloid Interface Sci.}
  \textbf{\bibinfo{volume}{89}}, \bibinfo{pages}{563} (\bibinfo{year}{1982}).

\bibitem[{\citenamefont{Gopinathan et~al.}(2002)\citenamefont{Gopinathan, Zhou,
  Coppersmith, Kadanoff, and Grier}}]{gopinathan02}
\bibinfo{author}{\bibfnamefont{A.}~\bibnamefont{Gopinathan}},
  \bibinfo{author}{\bibfnamefont{T.}~\bibnamefont{Zhou}},
  \bibinfo{author}{\bibfnamefont{S.~N.} \bibnamefont{Coppersmith}},
  \bibinfo{author}{\bibfnamefont{L.~P.} \bibnamefont{Kadanoff}},
  \bibnamefont{and} \bibinfo{author}{\bibfnamefont{D.~G.} \bibnamefont{Grier}},
  \bibinfo{journal}{Europhys. Lett.} \textbf{\bibinfo{volume}{57}},
  \bibinfo{pages}{451} (\bibinfo{year}{2002}).

\bibitem[{\citenamefont{Han and Grier}(2004)}]{han04}
\bibinfo{author}{\bibfnamefont{Y.}~\bibnamefont{Han}} \bibnamefont{and}
  \bibinfo{author}{\bibfnamefont{D.~G.} \bibnamefont{Grier}},
  \bibinfo{journal}{Phys. Rev. Lett.} \textbf{\bibinfo{volume}{92}},
  \bibinfo{pages}{148301} (\bibinfo{year}{2004}).

\bibitem[{\citenamefont{Han and Grier}(2005)}]{han05a}
\bibinfo{author}{\bibfnamefont{Y.}~\bibnamefont{Han}} \bibnamefont{and}
  \bibinfo{author}{\bibfnamefont{D.~G.} \bibnamefont{Grier}},
  \bibinfo{journal}{J. Chem. Phys.} \textbf{\bibinfo{volume}{122}},
  \bibinfo{pages}{064907} (\bibinfo{year}{2005}).

\bibitem[{\citenamefont{Squires and Brenner}(2000)}]{squires00}
\bibinfo{author}{\bibfnamefont{T.~M.}~\bibnamefont{Squires}} \bibnamefont{and}
  \bibinfo{author}{\bibfnamefont{M.~P.} \bibnamefont{Brenner}},
  \bibinfo{journal}{Phys. Rev. Lett.} \textbf{\bibinfo{volume}{85}},
  \bibinfo{pages}{4976} (\bibinfo{year}{2000}).

\bibitem[{\citenamefont{Brunner et~al.}(2002)\citenamefont{Brunner, Bechinger,
  Strepp, Lobaskin, and von Grunberg. H.~H.}}]{brunner02}
\bibinfo{author}{\bibfnamefont{M.}~\bibnamefont{Brunner}},
  \bibinfo{author}{\bibfnamefont{C.}~\bibnamefont{Bechinger}},
  \bibinfo{author}{\bibfnamefont{W.}~\bibnamefont{Strepp}},
  \bibinfo{author}{\bibfnamefont{V.}~\bibnamefont{Lobaskin}}, \bibnamefont{and}
  \bibinfo{author}{\bibnamefont{von Grunberg. H.~H.}},
  \bibinfo{journal}{Europhys. Lett.} \textbf{\bibinfo{volume}{58}},
  \bibinfo{pages}{926} (\bibinfo{year}{2002}).

\bibitem[{\citenamefont{Baumgartl et~al.}(2005)\citenamefont{Baumgartl,
  Arauz-Lara, and Bechinger}}]{baumgartl05}
\bibinfo{author}{\bibfnamefont{J.}~\bibnamefont{Baumgartl}},
  \bibinfo{author}{\bibfnamefont{J.~L.} \bibnamefont{Arauz-Lara}},
  \bibnamefont{and}
  \bibinfo{author}{\bibfnamefont{C.}~\bibnamefont{Bechinger}},
  \bibinfo{journal}{preprint}  (\bibinfo{year}{2005}).

\bibitem[{\citenamefont{Baumgartl and Bechinger}(2005)}]{baumgartl05a}
\bibinfo{author}{\bibfnamefont{J.}~\bibnamefont{Baumgartl}} \bibnamefont{and}
  \bibinfo{author}{\bibfnamefont{C.}~\bibnamefont{Bechinger}},
  \bibinfo{journal}{Europhys. Lett.} \textbf{\bibinfo{volume}{71}},
  \bibinfo{pages}{487} (\bibinfo{year}{2005}).

\bibitem[{\citenamefont{Crocker}(1996)}]{crocker96phd}
\bibinfo{author}{\bibfnamefont{J.~C.} \bibnamefont{Crocker}},
  \bibinfo{type}{Ph.d.}, \bibinfo{school}{University of Chicago}
  (\bibinfo{year}{1996}).

\bibitem[{\citenamefont{Cui et~al.}(2002)\citenamefont{Cui, Lin, Sharma, and
  Rice}}]{cui02}
\bibinfo{author}{\bibfnamefont{B.}~\bibnamefont{Cui}},
  \bibinfo{author}{\bibfnamefont{B.}~\bibnamefont{Lin}},
  \bibinfo{author}{\bibfnamefont{S.}~\bibnamefont{Sharma}}, \bibnamefont{and}
  \bibinfo{author}{\bibfnamefont{S.~A.} \bibnamefont{Rice}},
  \bibinfo{journal}{J. Chem. Phys.} \textbf{\bibinfo{volume}{116}},
  \bibinfo{pages}{3119} (\bibinfo{year}{2002}).

\bibitem[{\citenamefont{Crocker and Grier}(1996{\natexlab{b}})}]{crocker96}
\bibinfo{author}{\bibfnamefont{J.~C.} \bibnamefont{Crocker}} \bibnamefont{and}
  \bibinfo{author}{\bibfnamefont{D.~G.} \bibnamefont{Grier}},
  \bibinfo{journal}{J. Colloid Interface Sci.} \textbf{\bibinfo{volume}{179}},
  \bibinfo{pages}{298} (\bibinfo{year}{1996}{\natexlab{b}}).

\bibitem[{\citenamefont{Bohren and Huffman}(1983)}]{bohren83}
\bibinfo{author}{\bibfnamefont{C.~F.} \bibnamefont{Bohren}} \bibnamefont{and}
  \bibinfo{author}{\bibfnamefont{D.~R.} \bibnamefont{Huffman}},
  \emph{\bibinfo{title}{Absorption and Scattering of Light by Small Particles}}
  (\bibinfo{publisher}{Wiley Interscience}, \bibinfo{address}{New York},
  \bibinfo{year}{1983}).

\bibitem[{\citenamefont{Ovryn and Izen}(2000)}]{ovryn00}
\bibinfo{author}{\bibfnamefont{B.}~\bibnamefont{Ovryn}} \bibnamefont{and}
  \bibinfo{author}{\bibfnamefont{S.~H.} \bibnamefont{Izen}},
  \bibinfo{journal}{J. Opt. Soc. America} \textbf{\bibinfo{volume}{17}},
  \bibinfo{pages}{1202} (\bibinfo{year}{2000}).

\bibitem[{\citenamefont{Dufresne and Grier}(1998)}]{dufresne98}
\bibinfo{author}{\bibfnamefont{E.~R.} \bibnamefont{Dufresne}} \bibnamefont{and}
  \bibinfo{author}{\bibfnamefont{D.~G.} \bibnamefont{Grier}},
  \bibinfo{journal}{Rev. Sci. Instr.} \textbf{\bibinfo{volume}{69}},
  \bibinfo{pages}{1974} (\bibinfo{year}{1998}).

\bibitem[{\citenamefont{Polin et~al.}(2005)\citenamefont{Polin, Ladavac, Lee,
  Roichman, and Grier}}]{polin05}
\bibinfo{author}{\bibfnamefont{M.}~\bibnamefont{Polin}},
  \bibinfo{author}{\bibfnamefont{K.}~\bibnamefont{Ladavac}},
  \bibinfo{author}{\bibfnamefont{S.-H.} \bibnamefont{Lee}},
  \bibinfo{author}{\bibfnamefont{Y.}~\bibnamefont{Roichman}}, \bibnamefont{and}
  \bibinfo{author}{\bibfnamefont{D.~G.} \bibnamefont{Grier}},
  \bibinfo{journal}{Opt. Express} \textbf{\bibinfo{volume}{13}},
  \bibinfo{pages}{5831} (\bibinfo{year}{2005}).

\bibitem[{\citenamefont{Chan}(1977)}]{chan77}
\bibinfo{author}{\bibfnamefont{E.~M.} \bibnamefont{Chan}}, \bibinfo{journal}{J.
  Phys. C} \textbf{\bibinfo{volume}{10}}, \bibinfo{pages}{3477}
  (\bibinfo{year}{1977}).

\bibitem[{\citenamefont{Russel et~al.}(1989)\citenamefont{Russel, Saville, and
  Schowalter}}]{russel89}
\bibinfo{author}{\bibfnamefont{W.~B.} \bibnamefont{Russel}},
  \bibinfo{author}{\bibfnamefont{D.~A.} \bibnamefont{Saville}},
  \bibnamefont{and} \bibinfo{author}{\bibfnamefont{W.~R.}
  \bibnamefont{Schowalter}}, \emph{\bibinfo{title}{Colloidal Dispersions}},
  Cambridge Monographs on Mechanics and Applied Mathematics
  (\bibinfo{publisher}{Cambridge University Press},
  \bibinfo{address}{Cambridge}, \bibinfo{year}{1989}).

\bibitem[{\citenamefont{Russ et~al.}(2005)\citenamefont{Russ, Brunner,
  Bechinger, and Von~Grunberg}}]{russ05}
\bibinfo{author}{\bibfnamefont{C.}~\bibnamefont{Russ}},
  \bibinfo{author}{\bibfnamefont{M.}~\bibnamefont{Brunner}},
  \bibinfo{author}{\bibfnamefont{C.}~\bibnamefont{Bechinger}},
  \bibnamefont{and} \bibinfo{author}{\bibfnamefont{H.~H.}
  \bibnamefont{Von~Grunberg}}, \bibinfo{journal}{Europhys. Lett.}
  \textbf{\bibinfo{volume}{69}}, \bibinfo{pages}{468} (\bibinfo{year}{2005}).

\bibitem[{\citenamefont{Hastings}(1978)}]{hastings78}
\bibinfo{author}{\bibfnamefont{R.}~\bibnamefont{Hastings}},
  \bibinfo{journal}{J. Chem. Phys.} \textbf{\bibinfo{volume}{68}},
  \bibinfo{pages}{675} (\bibinfo{year}{1978}).

\bibitem[{\citenamefont{Lee and Fisher}(1997)}]{lee97}
\bibinfo{author}{\bibfnamefont{B.~P.} \bibnamefont{Lee}} \bibnamefont{and}
  \bibinfo{author}{\bibfnamefont{M.~E.} \bibnamefont{Fisher}},
  \bibinfo{journal}{Europhys. Lett.} \textbf{\bibinfo{volume}{39}},
  \bibinfo{pages}{611} (\bibinfo{year}{1997}).

\bibitem[{\citenamefont{Carbajal-Tinoco and
  Gonzalez-Mozuelos}(2002)}]{carbajaltinoco02}
\bibinfo{author}{\bibfnamefont{M.~D.} \bibnamefont{Carbajal-Tinoco}}
  \bibnamefont{and}
  \bibinfo{author}{\bibfnamefont{P.}~\bibnamefont{Gonzalez-Mozuelos}},
  \bibinfo{journal}{J. Chem. Phys.} \textbf{\bibinfo{volume}{117}},
  \bibinfo{pages}{2344} (\bibinfo{year}{2002}).

\bibitem[{\citenamefont{Goulding and Hansen}(1998)}]{goulding98}
\bibinfo{author}{\bibfnamefont{D.}~\bibnamefont{Goulding}} \bibnamefont{and}
  \bibinfo{author}{\bibfnamefont{J.-P.} \bibnamefont{Hansen}},
  \bibinfo{journal}{Mol. Phys.} \textbf{\bibinfo{volume}{95}},
  \bibinfo{pages}{649} (\bibinfo{year}{1998}).

\bibitem[{\citenamefont{Goulding and Hansen}(1999)}]{goulding99a}
\bibinfo{author}{\bibfnamefont{D.}~\bibnamefont{Goulding}} \bibnamefont{and}
  \bibinfo{author}{\bibfnamefont{J.-P.} \bibnamefont{Hansen}},
  \bibinfo{journal}{Europhys. Lett.} \textbf{\bibinfo{volume}{46}},
  \bibinfo{pages}{407} (\bibinfo{year}{1999}).

\bibitem[{\citenamefont{Mateescu}(2001)}]{mateescu01}
\bibinfo{author}{\bibfnamefont{E.~M.} \bibnamefont{Mateescu}},
  \bibinfo{journal}{Phys. Rev. E} \textbf{\bibinfo{volume}{64}},
  \bibinfo{pages}{013401} (\bibinfo{year}{2001}).

\bibitem[{\citenamefont{Trizac and Raimbault}(2001)}]{trizac01}
\bibinfo{author}{\bibfnamefont{E.}~\bibnamefont{Trizac}} \bibnamefont{and}
  \bibinfo{author}{\bibfnamefont{J.~L.} \bibnamefont{Raimbault}},
  \bibinfo{journal}{Phys. Rev. E} \textbf{\bibinfo{volume}{64}},
  \bibinfo{pages}{043401} (\bibinfo{year}{2001}).

\bibitem[{\citenamefont{Allahyarov et~al.}(1999)\citenamefont{Allahyarov,
  D'Amico, and L\"{o}wen}}]{allahyarov99}
\bibinfo{author}{\bibfnamefont{E.}~\bibnamefont{Allahyarov}},
  \bibinfo{author}{\bibfnamefont{I.}~\bibnamefont{D'Amico}}, \bibnamefont{and}
  \bibinfo{author}{\bibfnamefont{H.}~\bibnamefont{L\"{o}wen}},
  \bibinfo{journal}{Phys. Rev. E} \textbf{\bibinfo{volume}{60}},
  \bibinfo{pages}{3199} (\bibinfo{year}{1999}).

\bibitem[{\citenamefont{Grier and Han}(2004)}]{grier04f}
\bibinfo{author}{\bibfnamefont{D.~G.} \bibnamefont{Grier}} \bibnamefont{and}
  \bibinfo{author}{\bibfnamefont{Y.}~\bibnamefont{Han}}, \bibinfo{journal}{J.
  Phys.: Condens. Matt.} \textbf{\bibinfo{volume}{16}}, \bibinfo{pages}{S4145}
  (\bibinfo{year}{2004}).

\end{thebibliography}

\end{document}